\documentclass[final,5p,times,twocolumn]{elsarticle}

\usepackage{epsfig}
\usepackage{here}
\usepackage{amssymb}

\usepackage{lineno}

\usepackage{color}
\usepackage{amsmath}

\journal{Astroparticle Physics}

\newcommand{\km}{\ensuremath{\mathrm{km}}}

\newcommand{\GeV}{\ensuremath{\mathrm{GeV}}}

\begin{document}

\begin{frontmatter}



\title{Search for WIMP-$^{129}$Xe inelastic scattering with particle identification in XMASS-I}

\address{\rm\normalsize XMASS Collaboration$^*$}
\author[ICRR]{T.~Suzuki}
\author[ICRR,IPMU]{K.~Abe}
\author[ICRR,IPMU]{K.~Hiraide}
\author[ICRR,IPMU]{K.~Ichimura}
\author[ICRR,IPMU]{Y.~Kishimoto}
\author[ICRR,IPMU]{K.~Kobayashi}
\author[ICRR]{M.~Kobayashi}
\author[ICRR,IPMU]{S.~Moriyama}
\author[ICRR,IPMU]{M.~Nakahata}
\author[ICRR,IPMU]{H.~Ogawa\fnref{OgawaNow}}
\author[ICRR]{K.~Sato}
\author[ICRR,IPMU]{H.~Sekiya}
\author[ICRR,IPMU]{A.~Takeda}
\author[ICRR]{S.~Tasaka}
\author[ICRR,IPMU]{M.~Yamashita}
\author[ICRR,IPMU]{B.~S.~Yang\fnref{YangNow}}
\author[IBS]{N.~Y.~Kim}
\author[IBS]{Y.~D.~Kim}
\author[ISEE,KMI]{Y.~Itow}
\author[ISEE]{K.~Kanzawa}
\author[ISEE]{K.~Masuda}
\author[IPMU]{K.~Martens}
\author[IPMU]{Y.~Suzuki}
\author[IPMU]{B.~D.~Xu}
\author[Kobe]{K.~Miuchi}
\author[Kobe]{N.~Oka}
\author[Kobe,IPMU]{Y.~Takeuchi}
\author[KRISS,IBS]{Y.~H.~Kim}
\author[KRISS]{K.~B.~Lee}
\author[KRISS]{M.~K.~Lee}
\author[Miyagi]{Y.~Fukuda}
\author[Tokai1]{M.~Miyasaka}
\author[Tokai1]{K.~Nishijima}
\author[Tokushima]{K.~Fushimi}
\author[Tokushima]{G.~Kanzaki}
\author[YNU1]{S.~Nakamura}
\address[ICRR]{Kamioka Observatory, Institute for Cosmic Ray Research, the University of Tokyo, Higashi-Mozumi, Kamioka, Hida, Gifu, 506-1205, Japan}
\address[IBS]{Center for Underground Physics, Institute for Basic Science, 70 Yuseong-daero 1689-gil, Yuseong-gu, Daejeon, 305-811, South Korea}
\address[ISEE]{Institute for Space-Earth Environmental Research, Nagoya University, Nagoya, Aichi 464-8601, Japan}
\address[IPMU]{Kavli Institute for the Physics and Mathematics of the Universe (WPI), the University of Tokyo, Kashiwa, Chiba, 277-8582, Japan}
\address[KMI]{Kobayashi-Maskawa Institute for the Origin of Particles and the Universe, Nagoya University, Furo-cho, Chikusa-ku, Nagoya, Aichi, 464-8602, Japan}
\address[Kobe]{Department of Physics, Kobe University, Kobe, Hyogo 657-8501, Japan}
\address[KRISS]{Korea Research Institute of Standards and Science, Daejeon 305-340, South Korea}
\address[Miyagi]{Department of Physics, Miyagi University of Education, Sendai, Miyagi 980-0845, Japan}
\address[Tokai1]{Department of Physics, Tokai University, Hiratsuka, Kanagawa 259-1292, Japan}
\address[Tokushima]{Department of Physics, Tokushima University, 2-1 Minami Josanjimacho Tokushima city, Tokushima, 770-8506, Japan}
\address[YNU1]{Department of Physics, Faculty of Engineering, Yokohama National University, Yokohama, Kanagawa 240-8501, Japan}

\cortext[CollabEMail]{{\it E-mail address:} xmass.publications15@km.icrr.u-tokyo.ac.jp}
\fntext[OgawaNow]{Now at Department of Physics, College of Science and Technology, Nihon University, Kanda, Chiyoda-ku, Tokyo, 101-8308, Japan.}
\fntext[YangNow]{Now at Center for Axion and Precision Physics Research, Institute for Basic Science, Daejeon 34051, South Korea.}

\begin{abstract}
A search for Weakly Interacting Massive Particles (WIMPs) was conducted
with the single-phase liquid-xenon detector XMASS through inelastic scattering in which $^{129}$Xe nuclei were excited, using an exposure ($\rm 327\; kg \times 800.0 \; days$) 48 times larger than that of our previous study.
The inelastic excitation sensitivity was improved by detailed evaluation of background, event classification based on scintillation timing that distinguished $\gamma$-rays and $\beta$-rays, and simultaneous fitting of the energy spectra of $\gamma$-like and $\beta$-like samples. 
No evidence of a WIMP signal was found. 
Thus, we set the upper limits of the inelastic channel cross section at 90\% confidence level, for example, 
$4.1\times 10^{-39} \;{\rm cm^2}$ for a $200\; {\rm GeV}/c^2$ WIMP. 
This result provides the most stringent limits on the SD WIMP-neutron interaction
and is better by a factor of 7.7 at $200\;{\rm GeV}/c^2$ than the existing experimental limit.

\end{abstract}


\end{frontmatter}


%
%

\section{Introduction}

A considerable amount of evidence suggesting the existence of dark matter has been found through the optical observation and theoretical prediction of the rotational curve of galaxies, 
 gravitational lensing, etc \cite{pdg}.
Among the many theoretical candidates for dark matter, Weakly Interacting Massive Particles (WIMPs) are of particular interest in direct detection experiments.
If WIMPs exist, it is expected that their interaction with baryonic matter would be strong enough for nuclear recoils to be observed.
However, despite ongoing global efforts, neither direct nor indirect detection has yet been achieved. 

The interactions between WIMPs and nuclei should come in two types, Spin-Independent (SI) and Spin-Dependent (SD) interactions. 
SI interactions are often searched for via elastic scattering \cite{pdg}.
SD interactions are possible if WIMPs have non-zero spin. It allows for both elastic and inelastic scattering. 
The target nuclei should have effective nuclear spin.
Odd-mass number nuclei, e.g. $^{127}$I, $^{129}$Xe, and $^{131}$Xe satisfy that requirement and can be used for the SD search.
Because the $^{129}$Xe nucleus contains an unpaired neutron, we expect a large SD WIMP-neutron cross section in the shell model.
Its SD WIMP-proton cross section is smaller than its SD WIMP-neutron cross section by one order of magnitude because it only contains paired protons.
A couple of searches for SD interactions via WIMP-nucleon elastic scattering gave null results \cite{cdmsela,zeplinela,xenon10ela,xenon100ela,luxela,pandaela}. 
However, there is the difficulty of distinguishing between SD and SI interactions in elastic scattering. 
On the other hand, an observation of WIMP-nuclei inelastic scattering would be direct evidence of an SD interaction mechanism as well as that WIMPs have spin since 
nuclear excitation in inelastic scattering can be led only by SD interaction. 
Thus the search for inelastic scattering by WIMPs is important approach to the nature of SD interaction although the sensitivity of the search for WIMPs via inelastic channel is an order of magnitude worse than that via elastic channel \cite{baudis}.

In the past, searches for inelastic scattering were conducted using $^{127}$I \cite{fushimi} or $^{129}$Xe. Searches with $^{129}$Xe were first performed by the DAMA group in 1996 and 2000 \cite{dama96,dama00}. 
XMASS obtained a 90\% Confidence Level (CL) upper limit on the SD WIMP-neutron cross section at $4.2\times 10^{-38}\; {\rm cm^2}$ for a $50 \; {\rm GeV}/c^2$ WIMP with an exposure of $\rm 41\; kg\times 132.0\; days$ in 2014 \cite{uchida}. XENON100 published an upper limit of $3.3\times 10^{-38}\; {\rm cm^2}$ for a $100 \; {\rm GeV}/c^2$ WIMP with $\rm 34\; kg\times 224.6\; days$ exposure in 2017 \cite{xenon100}.

In this paper, an improved result for the search of inelastic scattering in XMASS is reported. An exposure of $\rm 327\; kg\times 800.0 \; days$ was accumulated and analyzed after the refurbishment of the XMASS detector \cite{fvana}. 
In addition to the increased exposure, an analysis update including detailed evaluation of background (BG) and particle identification improved the sensitivity.

\section{XMASS-I detector}

The XMASS-I detector is a single-phase detector containing 832 kg of liquid xenon (LXe) and located approximately 1,000 meters underground in the Kamioka mine (2,700 meter water equivalent) \cite{detector}.
The geometry of its sensitive volume is a pentakis-dodecahedron, with an inscribed radius of approximately $\rm 40\; cm$.
Scintillation light from the LXe in the sensitive volume is detected by 642 Hamamatsu R10789 Photo-Multiplier Tubes (PMTs), which have typical quantum efficiencies of $\sim 30\%$.
An outer shell of LXe shields the inner fiducial volume against the external $\gamma$-rays, particularly those originating from the PMTs. 
The photocathodes of these PMTs cover 62.4\% of the detector's inner surface.
Signals from the PMTs are recorded by CAEN V1751 ($\rm 10\; bit$, $\rm 1\; GHz$) waveform digitizers.

To shield against fast neutrons and external $\gamma$-rays, the detector is surrounded by a cylindrical water tank, the height and diameter of which are $\rm 10.5\; m$ and $\rm 10\; m$, respectively.
This water tank is also referred to as the Outer Detector (OD) and is used as an active muon veto. 
The OD is equiped with 72 Hamamatsu H3600 (20-inch) PMTs.

Detector calibrations using $^{241}$Am and $^{57}$Co $\gamma$-ray sources are performed for tuning the optical parameters of the detector Monte Carlo simulation (MC), e.g. the scattering length and absorption length.
The sources are aligned with the vertical ($z$) axis of the detector, and the $\gamma$-ray calibration data is recorded at 10 cm intervals from $z=-40\;{\rm cm}$ to $z=40\; {\rm cm}$ around the center of the detector.
The $\gamma$-ray calibration data is also used for determining the scintillation time profile for the $\beta$-rays' and $\gamma$-rays' events.
A $^{252}$Cf neutron source is used to determine the timing parameters for the Nuclear Recoil (NR) events \cite{ichimura}.
The neutron source was installed at the end of a pipe, which penetrates the water region of the OD and reaches the vacuum vessel that thermally isolates the detector from the water.
The scintillation efficiency of the detector can be calculated by combining the result of the calibration and the non-linear model of the efficiency discussed in \cite{doke}.
Since the visible energy for the same deposited energy varies depending on the particle, the electron-equivalent energy unit $\rm keV_{ee}$ is used to represent the event energies. 

\section{Expected signal}

An inelastic scattering event occurring in $^{129}$Xe will have a nuclear recoil and an emission of a 39.6 keV $\gamma$-ray from the nuclear excitation. 
The contribution to the scintillation signal from the NR depends on the velocity distribution of the WIMPs in the galaxy as well as the nuclear form factor for SD interactions.
The differential event rate per unit visible energy of the NR component is \cite{baudis}

\begin{equation}
\begin{split}
\frac{dR}{dE_{\rm NRvis}}&=\frac{dE_{\rm NR}}{d(\mathcal{L}_{\rm eff} E_{\rm NR})}\frac{dR}{dE_{\rm NR}} \\
&=\frac{dE_{\rm NR}}{d(\mathcal{L}_{\rm eff} E_{\rm NR})}\frac{\rho_\chi \sigma }{2M_\chi \mu^2}\int _{v_{\rm min}(E_{\rm NR})}^{v_{\rm max}}\frac{1}{v}\frac{dn}{dv}\;dv\; ,
\end{split}
\label{drde}
\end{equation}
where $R$ is the event rate per unit target mass and unit time; $E_{\rm NRvis}$ is the energy represented using the unit $\rm keV_{ee}$;
$E_{\rm NR}$ is the energy of the recoiling nucleus;
$\mathcal{L}_{\rm eff}=E_{\rm NRvis}(E_{\rm NR})/E_{\rm NR}$ as described in \cite{leffref};
$\rho_\chi$ is the mass density of WIMPs in the laboratory for which we use the customary value of $0.3 \;\GeV/c^2/{\rm cm}^{3}$ \cite{density};
$M_\chi$ is the mass of the WIMP;
$\mu$ is the reduced mass of the WIMP and the target nucleus;
and $\sigma$ is the cross section for inelastic scattering.
This cross section can be obtained from the WIMP-neutron cross section $\sigma_{\rm neutron}$ as:

\begin{equation}
\sigma = \frac{4}{3}\frac{\pi}{2J+1}\left( \frac{\mu}{\mu_{\rm nucleon}}\right)^2S(E_{\rm NR})\:\sigma_{\rm neutron}\; ,
\end{equation}
where $J=1/2$ is the ground state spin of the $^{129}$Xe nucleus; $\mu_{\rm nucleon}$ is the reduced mass of the WIMP-nucleon system, and $S(E_{\rm NR})$ is the structure factor. We used ``$S_n(u)\; {\rm 1b+2b\; inelastic}$'' defined in \cite{baudis} as $S(E_{\rm NR})$. $v_{\rm min}(E_{\rm NR})$ is the minimum velocity of the WIMP needed to induce inelastic scattering with $E_{\rm NR}$;
$v_{\rm max}$ is the maximum velocity of WIMPs in the Earth's vicinity (544 km/s) \cite{escv},
and $dn/dv$ is the velocity distribution of the WIMPs.
WIMP velocities in the galaxy are assumed to follow a Gaussian distribution which is truncated at $v_{\rm max}$ and has a thermal speed of $220\;{\rm km/s}$ \cite{thv}.
Earth's velocity is assumed to be $232\; \km/{\rm s}$ \cite{earthv}.
$v_{\rm min}$ is evaluated to be

\begin{equation}
v_{\rm min}=v_{\rm min}^0+\frac{v_{\rm thr}^2}{4v_{\rm min}^0}\; ,
\end{equation}
where

\begin{equation}
v_{\rm min}^0=\sqrt{\frac{M_TE_{\rm NR}}{2\mu^2}},\;\;\; v_{\rm thr}^2=\frac{2\Delta E}{\mu}.
\end{equation}
Here, $M_T$ is the mass of target nucleus, and $\Delta E=39.58\;{\rm keV}$ is the energy of the $^{129}$Xe excited state.

\begin{figure}[t]
\begin{center}
\includegraphics[clip,width=9.0cm]{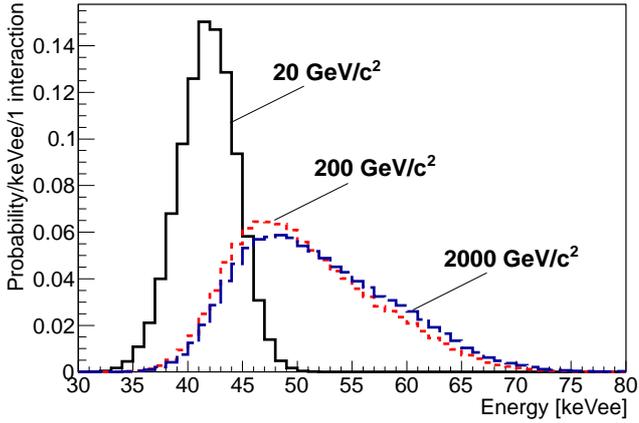}
\caption{Simulated energy spectra of the inelastic scattering events for 20 (solid), 200 (dotted), and 2000 (dashed) GeV/$c^2$ WIMPs. 
}
\label{spec200}
\end{center}
\end{figure}

MC was used to simulate the energy spectrum of the inelastic WIMP-nucleus collisions and BG spectra.
In the simulation, the recoil nucleus and de-excitation $\gamma$-ray are generated at the same time and position, 
since the  lifetime of the excited $^{129}$Xe is short enough ($< 1\;{\rm ns}$) to be ignored. 
The recoil energy distribution of the nucleus is based on $dR/dE_{\rm NR}$ in Eq. (\ref{drde}). 
The directions of the generated particles are isotropic, and the event vertices are uniformly distributed in the detector. 
Figure \ref{spec200} shows the simulated energy spectra for the inelastic scattering of 20, 200, and 2000 GeV/$c^2$ WIMPs. The NR component is more relevant for large mass WIMPs, and therefore they tend to have spectra with long tails to high energy.

\begin{figure}[t]
\begin{center}
\includegraphics[clip,width=9.0cm]{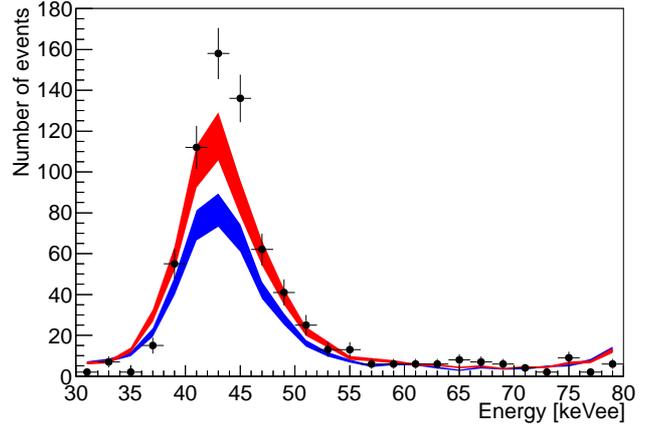}
\caption{Energy spectrum obtained from neutron calibration using a $^{252}$Cf source (black points) in the XMASS-I detector \cite{ichimura}. 
The blue and the red shaded spectra are from our MC with different cross section libraries, ENDF-B/VII \cite{endf71,endf72} and G4NDL3.13 based on ENDF/B-VI, respectively. 
Widths of these spectra represent the $\pm 10\%$ uncertainty of the neutron tagging efficiency \cite{ichimura}. }
\label{ncalib}
\end{center}
\end{figure}

\begin{figure*}[t]
\begin{center}
\includegraphics[clip,width=9.0cm]{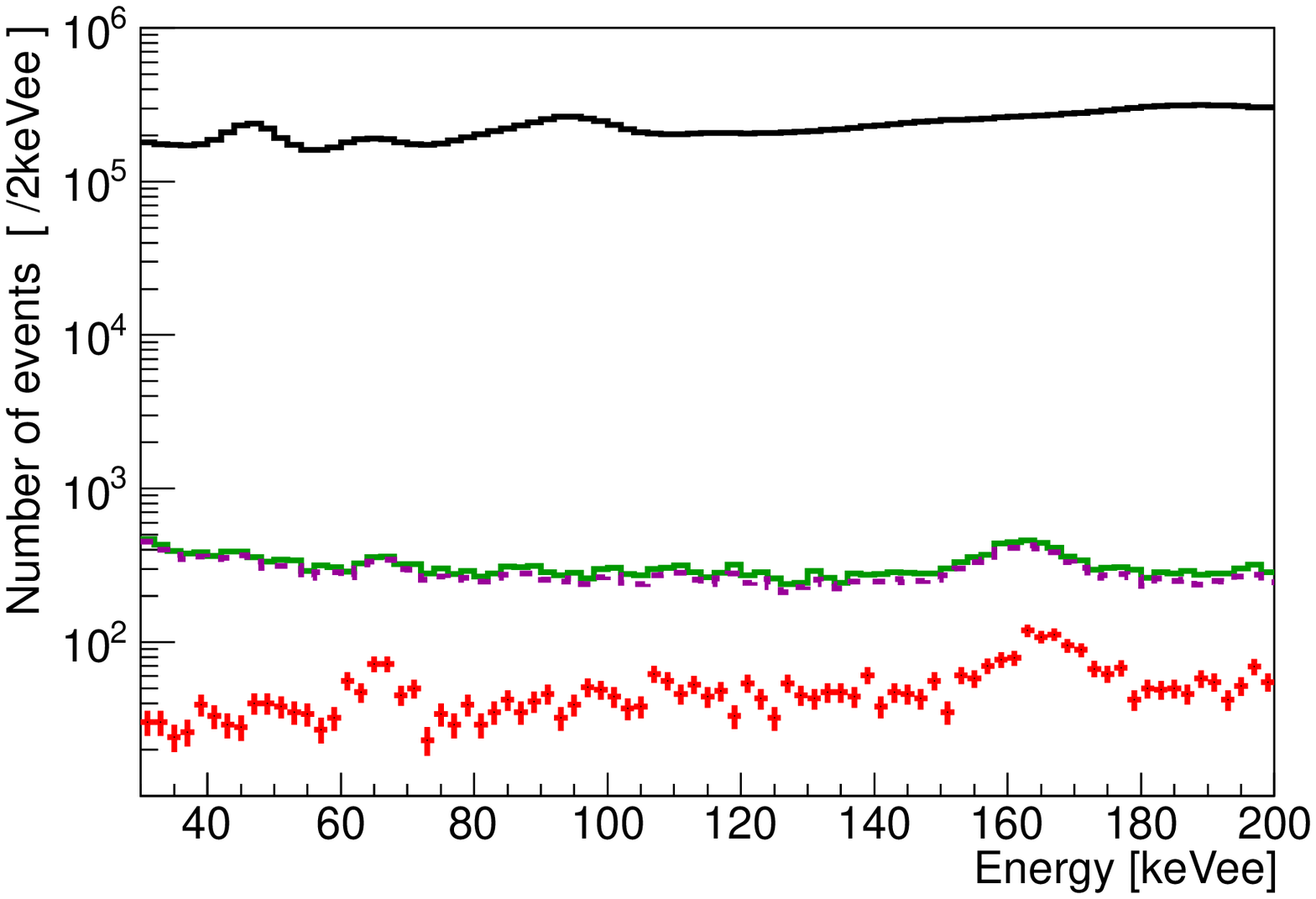}
\includegraphics[clip,width=9.0cm]{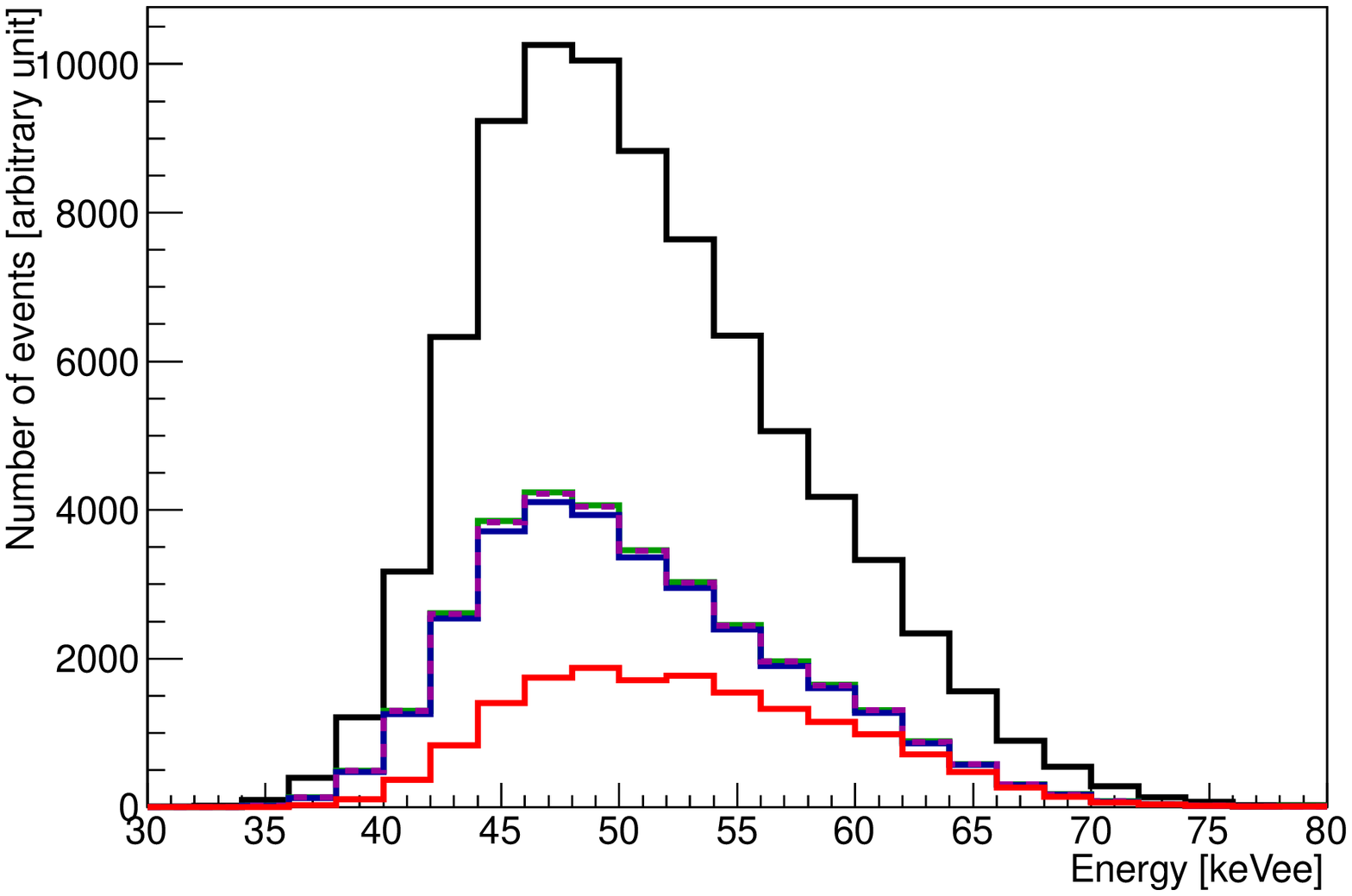}
\caption{ Result of events classification for the data (left) and $200\; {\rm GeV}/c^2$ WIMP MC (right). Histograms after pre-selection (black solid), fiducial volume cut (green solid), $^{214}$Bi rejection (magenta dashed), and $\beta$-like event rejection using $\beta$CL (red point/solid) are shown. 
  The left-hand figure depicts the entire energy region used druing the analysis 
($\rm 30\textendash 200\;  keV_{ee}$),
while the right-hand figure shows a magnified view of the $\rm 30\textendash 80\;  keV_{ee}$ region to make it easy to observe the WIMP signal.}
\label{class}
\end{center}
\end{figure*}

As a reference, the neutron inelastic scattering energy spectrum is presented in Figure \ref{ncalib}. 
These data were acquired from neutron calibration using a $^{252}$Cf source. 
Only the pre-selection cut was used in this figure (See Section \ref{refdataset}).
The fiducial volume cut was not used because a large fraction of neutron events occured outside the fiducial volume.  
The inelastic scattering peak was seen around $\rm 45\; keV_{ee}$. 

\section{Data and event classification}

\label{refdataset}
The data used for the analysis was collected between November 20, 2013 and July 20, 2016.
The detector was operated stably throughout the measurement period, during which the pressure above the LXe target was an absolute $\rm 0.162\textendash 0.164\; MPa$, and the temperature of the LXe was $\rm 172.6\textendash 173.0\; K$. 
The data taken within the ten days directly after the neutron calibrations was not used to reduce BG from activated Xe nuclei.
The data was divided into four periods, 1--4. 
The event rate due to neutron-activated xenon isotopes was relatively high during period 1 because of the following reasons:
\begin{enumerate}
\item Period 1 began only two weeks after the LXe was filled into the detector in the water shield.
\item Two neutron calibrations were performed during this period.
 \end{enumerate}
Period 2 started after these isotopes decayed and disappeared. Compared to period 1, the activities of $^{\rm 131m}$Xe and $^{133}$Xe decreased by factors of 4.3 and 1.3, respectively. 
A continuous gas circulation was started at the beginning of period 3. In the circulation, xenon gas extracted from the LXe was passed through a hot getter before being condensed into liquid. 
Before the start of period 4, we recovered the xenon from the detector in liquid phase to an external reservoir. Then we filled again the detector after purification by the hot getter. 
This procedure enabled us to remove potential non-volatile impurities from the detector.

In pre-selection, events stemming from the after pulses of PMTs caused by previous events were removed by choosing the events 
whose elapsed time from the previous inner-detector event ($dT_{\rm pre}$) was longer than $\rm 10\; ms$ and whose standard deviation of all the hit timings in the event was less than $\rm 100\; ns$.
The $dT_{\rm pre}$ requirement produces a dead time which corresponds to 3.0\% of the total livetime.

The event vertex was then reconstructed from the light distribution in the detector recorded by the PMTs \cite{detector}. 
The events whose vertices were reconstructed to be inside the fiducial volume were selected. 
In this analysis, the fiducial volume is a sphere with a radius of $\rm 30\; cm$ from the detector center. 
The total LXe in this fiducial volume is $\rm 327\; kg$ and contains $\rm 86\; kg$ of $^{129}$Xe.

The abundance of $^{222}$Rn progeny, which is a major source of BG, was estimated from the events in the fiducial volume. 
$^{214}$Bi events were tagged by looking for coincidences compatible with the $^{214}$Bi-$^{214}$Po decay sequence.
The time to the next event ($dT_{\rm post}$) was used to identify candidates. Since the half-life of $^{214}$Po is $164\;\mu{\rm s}$, 99.6\% of all $^{214}$Bi events can be tagged by selecting events with $0.015\;{\rm ms}<dT_{\rm post}< 1\; {\rm ms}$.
These tagged and non-tagged events will be referred to as the $^{214}$Bi and non-$^{214}$Bi samples, respectively.
0.4\% of non-Bi events were misplaced within the $^{214}$Bi sample.
This allowed for the Bi and Po concentration in the LXe to be estimated.

After $^{214}$Bi tagging, $\alpha$-events from the detector surface were eliminated from the non-$^{214}$Bi sample by choosing events whose scintillation time constant of the summed up PMT waveform was longer than $\rm 30\; ns$. 
 The decay time was obtained by fitting the data with an exponential function.
 By this selection, almost all $\alpha$-events were eliminated in the energy range above $30\; {\rm keV_{ee}}$, 
 while 97\% of the inelastic scattering events by a $200\;{\rm GeV}/c^2$ WIMPs remain.  
 The $\alpha$-events eliminated by this selection were considered to be produced outside the sensitive volume. 
 Since only a small fraction of scintillation photons was detected through small gaps around the PMTs, 
 they could be identified as low energy events. 
 Photons tend to cut through the fiducial volume and be detected also by the PMTs on the opposite side, not only by the PMTs around actual event location. 
 In such case, the events have some probability to be reconstructed inside the fiducial volume wrongly. This selection is effective for removing such background events.

The samples after the $\alpha$-event elimination were separated into $\beta$-depleted and $\beta$-enriched samples. 
This separation was performed with a particle identification technique based on the different LXe scintillation time profiles.
The time constant of scintillation from a $\beta$-ray becomes longer as the energy becomes larger \cite{takiya}. 
Since a $\gamma$-ray is converted into lower energy electrons in LXe, its time constant is shorter than that of a $\beta$-ray.
The scintillation light from the NR has a shorter time constant than that of a $\beta$-ray and a $\gamma$-ray since its ionization density is higher and the ion-electron pairs recombine faster \cite{ichimura}.
Since inelastic scattering has contributions from both NR and a $\gamma$-ray, it has a shorter time constant than a pure $\beta$-ray event.
Thus, in this process, the $\gamma$-ray events and the inelastic scattering events are expected to be preferentially sorted into the $\beta$-depleted sample. 
$\beta$CL, which represents the $p$-value of an event being a $\beta$-ray, is calculated using the cumulative distribution function (CDF) of a $\beta$-ray's scintillation timings \cite{pvalue,2nuecec}: 

\begin{equation}
\beta{\rm CL}=P\sum_{i=0}^{n-1}\frac{(-\ln P)^i}{i!}\;\;\;\; \left( P=\prod_{i=0}^{n-1}{\rm CDF}_\beta(E_{\rm evt},t_i) \right)\; ,
\label{bcl}
\end{equation}
where $n$ is the number of detected PMT pulses; $t_i$ is the timing of $i$-th pulse; $E_{\rm evt}$ is the event energy, and ${\rm CDF}_\beta (E_{\rm evt},t)$ is the CDF for finding a pulse at time $t$ in a $\beta$-event of energy $E_{\rm evt}$. 
${\rm CDF}_\beta (E_{\rm evt},t)$ was evaluated using the tagged $^{214}$Bi events. 
The evaluation was done with 1 ns timing bins and $5 \; {\rm keV_{ee}}$ energy and linear interpolation between bin centers. 
Theoretically, $\beta$CL distributes uniformly from 0 to 1 for $\beta$-ray events and for particles whose decay time is shorter than that of $\beta$-rays (such as $\gamma$-ray and NR), a peak appears near 0. Thus, $\gamma$-ray and inelastic scattering (NR together with a $\gamma$-ray) events are discriminated from $\beta$-ray events by $\beta$CL.
The probabilities that $\beta$-ray, $\gamma$-ray, and inelastic scattering are classified as $\beta$-depleted samples are referred to as $\beta$-ray misidentification probability ($\beta$ mis-ID), $\gamma$ efficiency, and signal efficiency, respectively. 
By setting a constant $\beta$CL threshold for event classification ($\beta{\rm CL_{th}}$) for all the energy region, the reduction ratio for $\beta$-rays becomes constant.  
On the other hand, 
since the contribution of NR component varies with the WIMP mass,
we set the $\beta{\rm CL_{th}}$ depending on the WIMP mass (e.g. $\beta{\rm CL_{th}}=0.06$ for a $200\;{\rm GeV}/c^2$ WIMP search). 
The $\beta{\rm CL_{th}}$ was optimized using MC so that $S/\sqrt{B}$ (the improvement factor of the significance of the signal) is maximized, where $S$ and $B$ are the signal efficiency and $\beta$ mis-ID, respectively. 

\begin{table}[t]
\begin{center}
\label{dataperiod1} 
\caption{Summary of the systematic uncertainty for each item. 
The threshold of $\beta$CL depends on the WIMP mass.
$\beta$CL-related uncertainties are for the $200\; {\rm GeV}/c^2$ WIMP search. 
}
  \begin{tabular}{lc}\hline \hline
    	& Fractional uncertainty  \\ 
    Item	& for each item  \\ \hline
 Energy scale& $ \pm 2 \%$	     \\ 
 Fiducial volume&  $^{+3.2}_{-4.0} \%$	     \\ 
 Thermal neutron flux& $ \pm 27 \%$	     \\ 
 $^{85}$Kr abundance in LXe & $ \pm 23  \%$	     \\ 
 $^{238}$U abundance in PMT & $ \pm 9.4  \%$	     \\ 
 $^{232}$Th abundance in PMT & $ \pm 24  \%$	     \\ 
 $^{60}$Co abundance in PMT & $ \pm 11  \%$	     \\ 
 $^{40}$K abundance in PMT & $ \pm 17  \%$	     \\ \hline
 $\beta$ mis-ID &   $  \pm 34 \%\;$  	\\
    $\gamma$ efficiency &  $ \pm 8.2 \%$  	     \\ 
Signal efficiency & $ \pm 8.5 \% $   \\
\hline \hline
  \end{tabular}
 \label{syserr} 
  \end{center}
\end{table}

\begin{figure}[t]
\begin{center}
\includegraphics[clip,width=9.0cm]{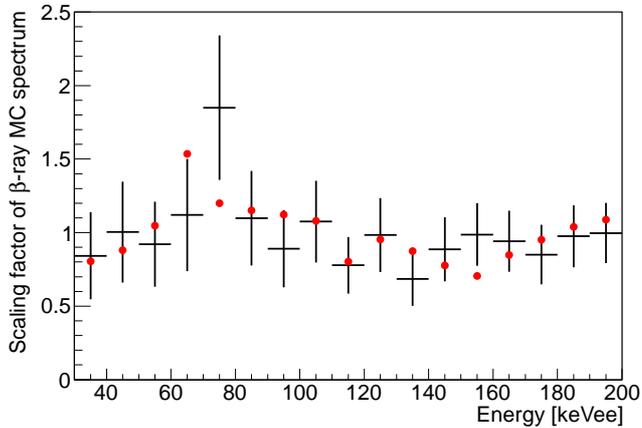}
\caption{Scaling factor of $\beta$-ray MC for the correction of $\beta$ mis-ID in a $200\;{\rm GeV}/c^2$ WIMP search. 
Black points are the means and errors evaluated using the difference between the data and MC results. 
Event rate of $\beta$-ray MC was scaled by a factor ``${\rm (mean)}+(p_l^{\rm const}-1)({\rm error})$''. The red points show the scale factor obtained by best fit.
}
\label{beta200}
\end{center}
\end{figure}

\begin{figure*}[t]
\begin{center}
\includegraphics[clip,width=15.0cm]{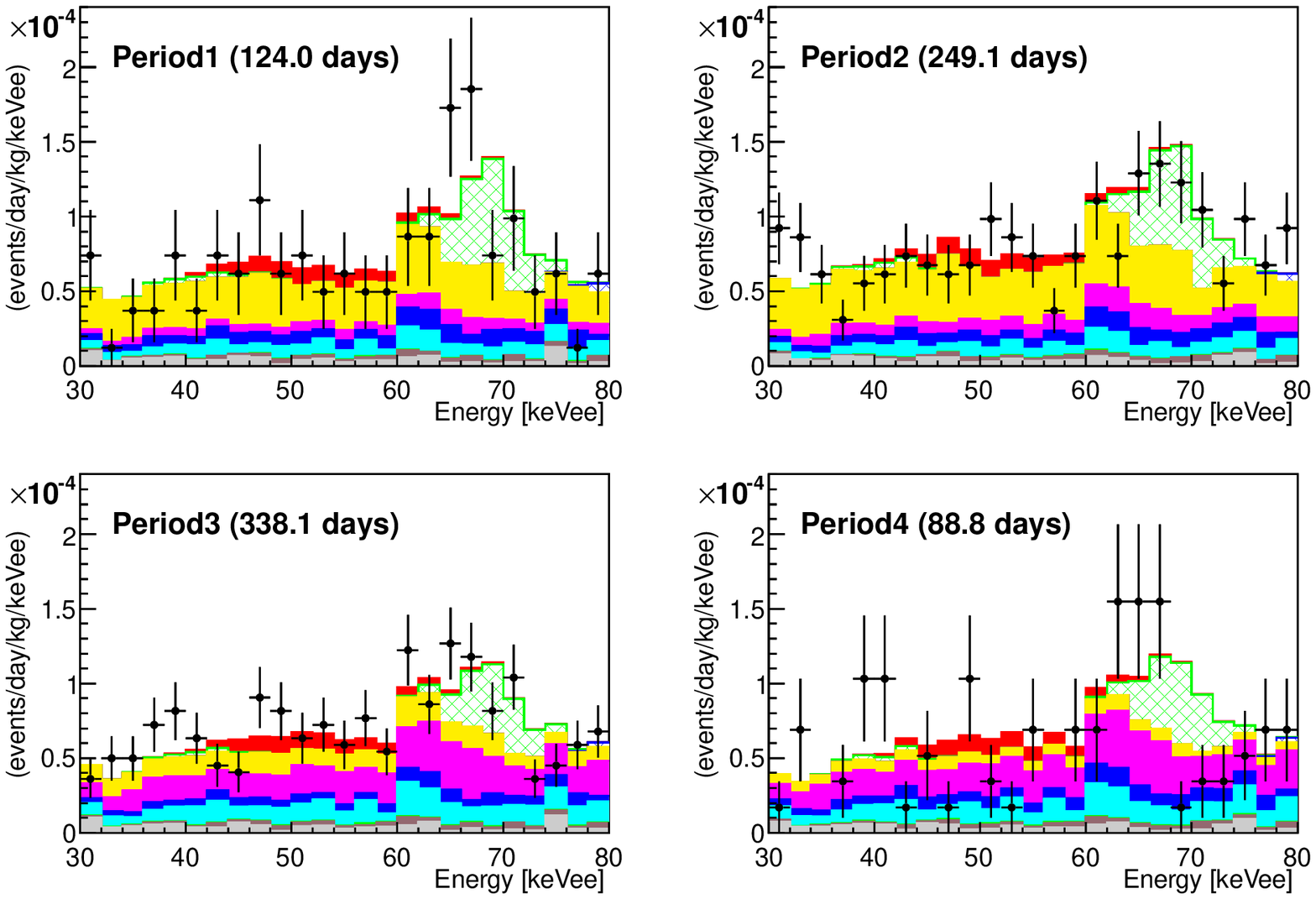}
\caption{$\beta$-depleted spectra with the 200 GeV/$c^2$ WIMP 90\% CL upper limit cross section. The observed data is shown as black points with error bars over the MC histograms. WIMP (red filled), $^{125}$I (green hatched), $^{14}$C (orange filled), $^{39}$Ar (magenta filled), $^{85}$Kr (blue filled), $^{214}$Pb (cyan filled), $^{136}$Xe (brown filled), and external $\gamma$-rays (gray filled) are shown as stacking histograms. Here, we show a magnified view of the $\rm 30\textendash 80\; keV_{ee}$ region to make it easy to observe the  WIMP signal.}
\label{zoom90}
\end{center}
\end{figure*}

The data and MC WIMP spectra during and after these treatments are shown in Figure \ref{class}. 
The number of events in the data decreased by two orders of magnitude after applying the fiducial volume cut, while the signal efficiency around the peak of the signal was remained at about 41\% of the events in full volume.
The difference in the number of events between before and after $^{214}$Bi reduction was taken to be a direct measure of the $^{214}$Bi event number. This number gives a constraint for the abundance of $^{214}$Pb, since both $^{214}$Bi and $^{214}$Pb are the progeny of $^{222}$Rn.
The $\beta$CL classification reduced the number of events by about one order of magnitude in the signal region.
For a $200\;{\rm GeV}/c^2$ WIMP, 
the signal efficiency in the fiducial volume region (the retained WIMP event ratio of before to after applying $^{214}$Bi, $\alpha$, and $\beta$-ray events reduction) is approximately 51\%. 
The $\beta$-ray events classified as $\beta$-depleted was typically about 10\% for a $200\;{\rm GeV}/c^2$ WIMP search.
To check the validity of the signal efficiency evaluation,
we applied the same event selection process, except for the fiducial
volume cut, to the neutron calibration and MC data in Figure \ref{ncalib} and 
found that their signal efficiencies were consistent.

\section{Energy spectrum fitting}

\label{secfit}

\renewcommand{\topfraction}{.99}
\renewcommand{\textfraction}{.01}
\setcounter{topnumber}{3}
\begin{figure}[t]
\begin{center}
\begin{minipage}{1.0\linewidth}
\includegraphics[clip,width=9cm]{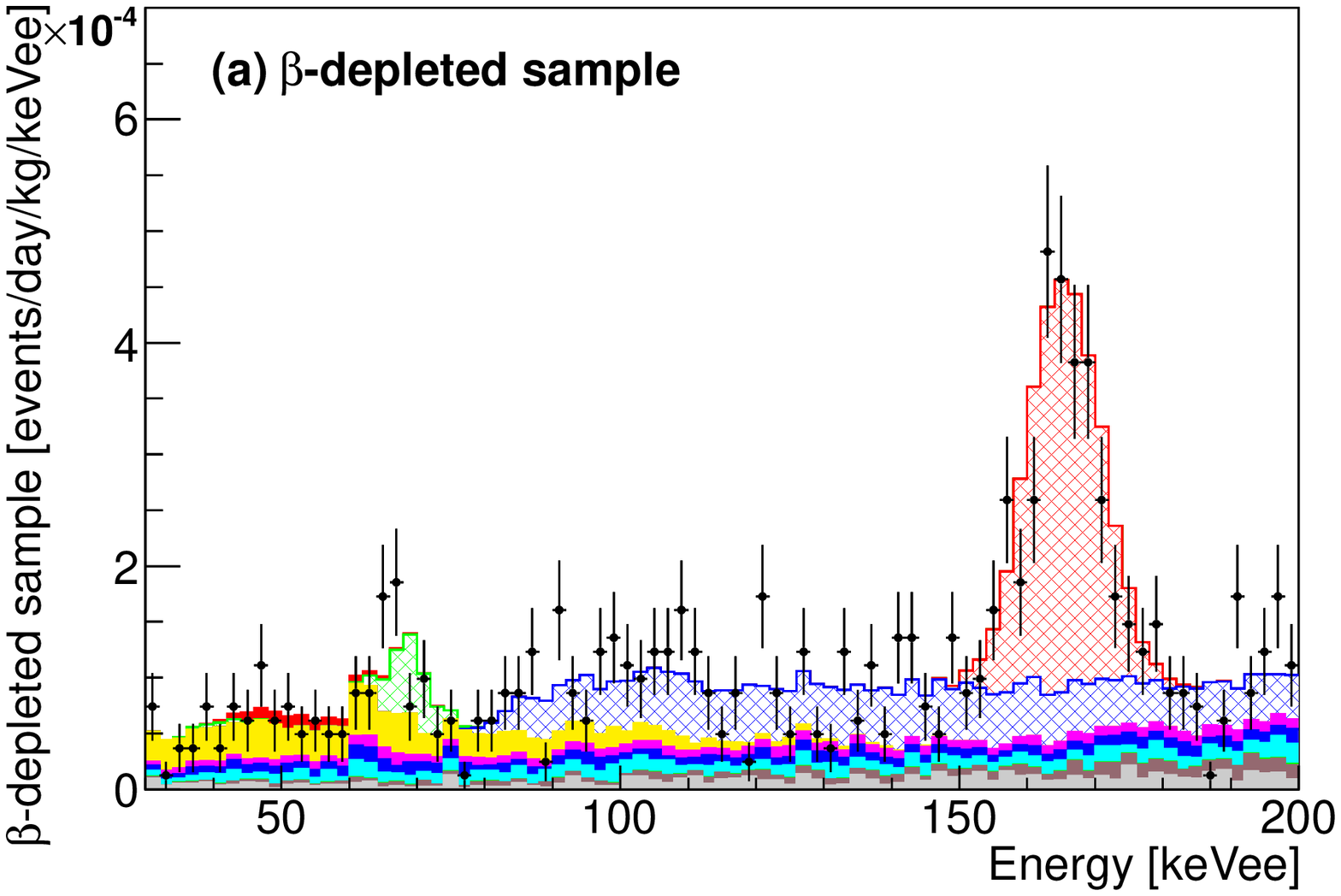}
\end{minipage}\\ 
\begin{minipage}{1.0\linewidth}
\includegraphics[clip,width=9cm]{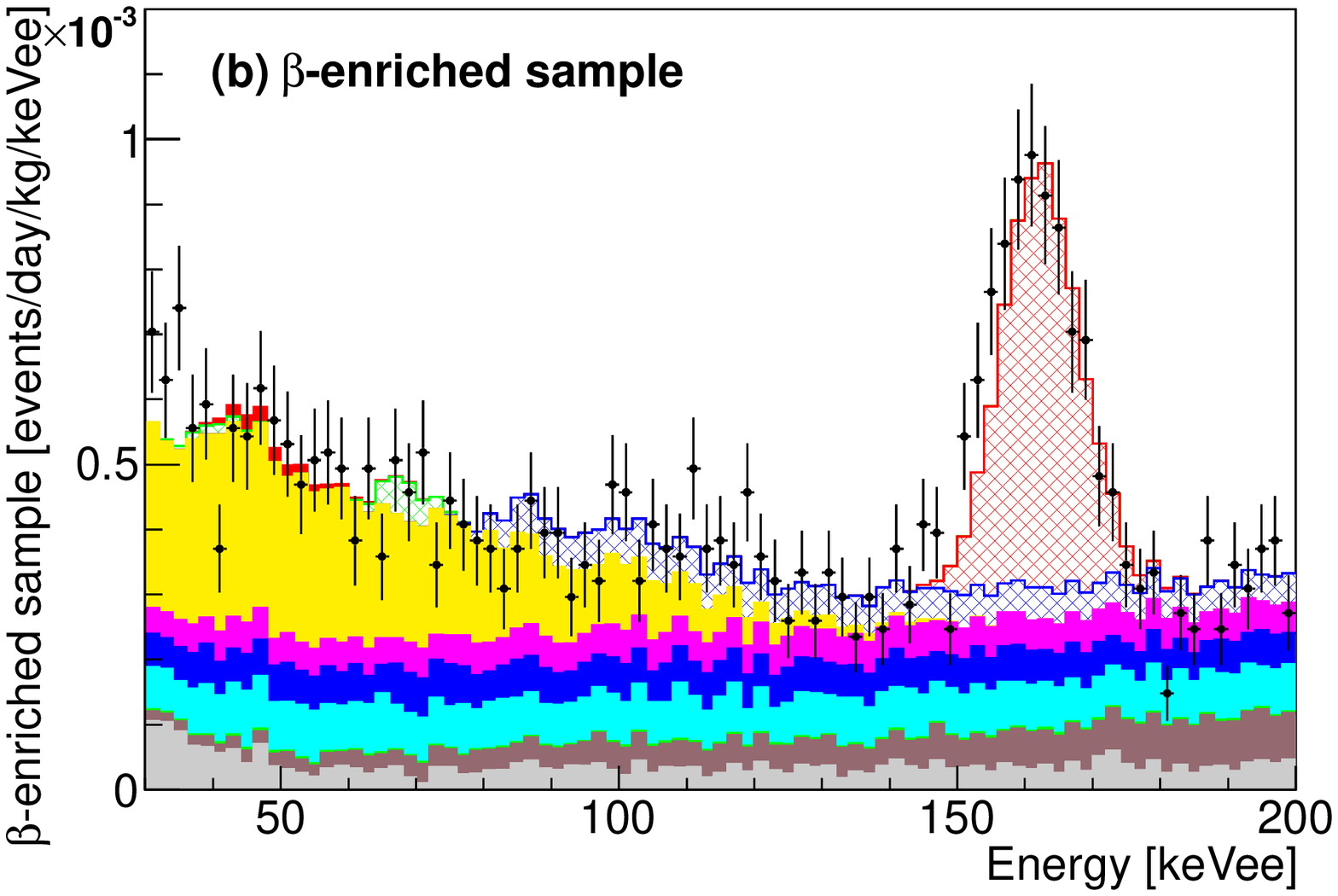}\\
\end{minipage}\\ 
\begin{minipage}{1.0\linewidth}
\includegraphics[clip,width=9cm]{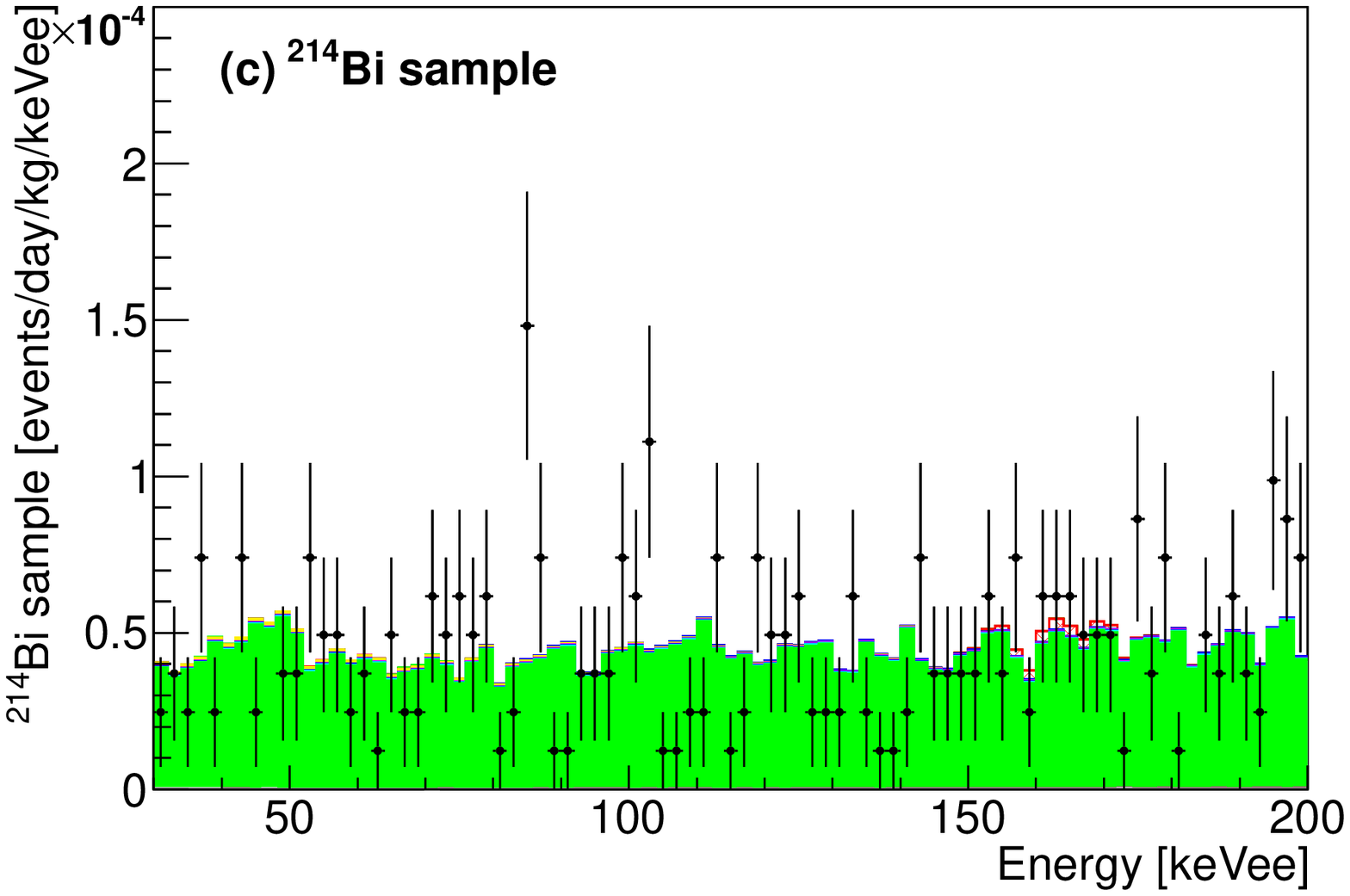}
\end{minipage}
\caption{Energy spectra for period 1 of $200\;{\rm GeV}/c^2$ WIMP (90\% CL upper limit). $\beta$-depleted, $\beta$-enriched, and $^{214}$Bi samples are shown in (a), (b), and (c), respectively. The definition of the color and hatch of histograms are the same as in Figure \ref{zoom90}. $^{\rm 131m}$Xe (red hatched), $^{133}$Xe (blue hatched), and $^{214}$Bi (green filled) are also shown.}
\label{allsample}
\end{center}
\end{figure}

In the previous section, the events were classified into three samples:
a $\beta$-depleted sample, a $\beta$-enriched sample, and a $^{214}$Bi sample.
By fitting the energy spectra of the $\beta$-depleted, the $\beta$-enriched, and the $^{214}$Bi samples simultaneously, we evaluated the amount of inelastic WIMP scattering that was compatible with our data. This fitting also determined the abundance of BG.
The activities of BG were estimated by the fit of the energy range from 30 to 200 $\rm keV_{ee}$. The width of the energy bins for the fit was 2 $\rm keV_{ee}$. 
The $\chi^2$ for the fit is defined as:

\begin{equation}
\begin{split}
\chi^2=&-2\ln L \\
      =&2\sum_{i=1}^{N_{\rm sample}}\sum_{j=1}^{N_{\rm period}}\sum_{k=1}^{N_{\rm bin}}
  \Biggl[ n^{\rm exp}_{ijk}(\{p^{\rm const}_l\},\{p^{\rm free}_m\})  \\
  & -n_{ijk}^{\rm data}+n^{\rm data}_{ijk}\ln \frac{n^{\rm exp}_{ijk}(\{p^{\rm const}_l\},\{p^{\rm free}_m\})}{n^{\rm data}_{ijk}} \Biggr] + \sum_{l=1}^{N_{\rm sys}}\frac{(1-p_l^{\rm const})^2}{\sigma _l ^2} \; ,
\end{split}
\label{chisquare}
\end{equation}
where $n_{ijk}^{\rm exp}$ is the total number of events including all BG MC and WIMP MC, and $n_{ijk}^{\rm data}$ is the number of events of the data. 
WIMP MC histogram was scaled by $\sigma_{\rm neutron}$.
Indices ``$i$'', ``$j$'', and ``$k$'' mean $i$-th sample, $j$-th period, and $k$-th energy bin, respectively. 
Here, $N_{\rm sample}=3$, $N_{\rm period}=4$ and $N_{\rm bin}=85$.
$p^{\rm const}_l\;(l=1,2,\cdots,N_{\rm sys})$ and $p^{\rm free}_m$ are scaling parameters for the constrained parameters and free parameters, respectively, which are described in detail below. 
The systematic uncertainty $\sigma_l$ is the $1\sigma$ constraint for $p^{\rm const}_l$.

The components of the BG are the radioisotopes (RIs) in LXe ($^{14}$C, $^{39}$Ar, $^{85}$Kr, $^{136}$Xe, $^{214}$Pb, and $^{214}$Bi),
the RIs in the PMTs ($^{238}$U, $^{232}$Th, $^{60}$Co, and $^{40}$K), 
RIs generated by thermal neutrons ($^{125}$I, $^{\rm 131m}$Xe, and $^{133}$Xe),
and the RIs in the detector's structure. 
For $^{14}$C and $^{39}$Ar, the constraints were not given and their abundances were determined by the fitting. The constraints for $^{85}$Kr and RIs in the PMTs were obtained by the BG study in XMASS \cite{fvana}.  The constraint for $^{125}$I was found using the result of a thermal neutron flux measurement in \cite{otani,minamino}. 
The constraint for the activities of $^{136}$Xe was given by KamLAND-Zen \cite{kamland}. 
The $^{214}$Bi sample gives constraints for the abundance of the daughters of $^{222}$Rn ($^{214}$Bi and $^{214}$Pb).
The impact of the RIs in the detector structure was found to be negligible.

The systematic uncertainty for the energy scale was evaluated by comparing the data and MC of $\rm ^{241}Am$'s 59.5 keV $\gamma$ and $\rm ^{57}Co$'s 122 keV $\gamma$. 
For the fiducial volume, the distributions of the reconstructed positions were compared between the $\rm ^{241}Am$ calibration data and MC at $z=30 \;{\rm cm}$.
Details of the evaluation of RIs' constraints and uncertainties for energy scale and fiducial volume are discussed in \cite{2nuecec}
and summarized in Table \ref{syserr}. Additional systematic uncertainties of the $\beta$CL-related values, i.e. the $\beta$ mis-ID, $\gamma$ efficiency, and signal efficiency defined in Section \ref{refdataset}, are discussed as follows and are also summarized in Table \ref{syserr}. 
MC histograms for each type of particle were scaled by using $\beta$CL-related uncertainties as the constraint. This is for the compensation of the discrepancy of $\beta$CL values between the data and MC.
\begin{enumerate}
\item {\it $\beta$ mis-ID}: The uncertainty of the $\beta$ mis-ID was obtained by comparing the data and MC of $^{214}$Bi.
Since the $\beta$-ray spectrum is continuous and covers the relevant energy region, this uncertainty was evaluated along with its energy-dependency.
The energy region from 30 to 200 $\rm keV_{ee}$ was divided into 17 bins and the difference of the probability that a $\beta$-ray event is classified into the $\beta$-depleted sample was compared between data and MC.
To correct the difference of $\beta$ mis-ID between the data and MC, $\beta$-ray BG MC histograms were scaled energy-dependently in the fitting using the $\beta$ mis-ID ratio between the data and MC. The scaling factor for a $200 \; {\rm GeV}/c^2$ WIMP search is shown in Figure \ref{beta200}.

\item {\it $\gamma$ efficiency}: The uncertainty in the efficiency of $\gamma$-ray retention was obtained using 59 keV and 122 keV $\gamma$-rays, again comparing the data and MC.
Here the evaluation was done independent of energy.
\item {\it Signal efficiency}: The uncertainty of the signal efficiency was evaluated by changing the timing parameters relevant for NR to their $\pm 1\sigma$ uncertainty range boundaries.
The change of the signal efficiency and this change in the NR timing parameters were used to evaluate the systematic uncertainty.
The relevant timing parameter values were obtained from the $^{252}$Cf calibration \cite{ichimura}.
\end{enumerate}

\section{Results and discussion}

The energy spectra of $\beta$-depleted, $\beta$-enriched, and $^{214}$Bi samples were fitted with the WIMP + BG spectra, where the WIMP mass was scanned between $20\; {\rm GeV}/c^2$ and $10\; {\rm TeV}/c^2$.
In the fitting, the BG abundances were determined for a given WIMP's cross section and mass.
The best fit cross section is defined by the minimum chi-square. The best fit cross section was  $7.0\times 10^{-40}\; {\rm cm^2}$ with $\chi^2/{\rm ndf}=1129/999$ for the $200\; {\rm GeV}/c^2$ WIMP. 
The minimum chi-square has no significant difference (within 1 $\sigma$) from that of the fitting without the WIMP signal at any WIMP mass.
Since no significant signal was found, the 90\% CL upper limit on the SD WIMP-neutron cross section was derived. 
To this end, the likelihood distribution $L(\sigma_{\rm neutron})$ for the cross section, i.e. the probability distribution of the cross section for the given experimental result, was evaluated:

\begin{equation}
  L(\sigma_{\rm neutron})=\exp{\left(-\frac{\chi ^2(\sigma_{\rm neutron})-\chi ^2(\sigma_{\rm min})}{2} \right)}
\label{ldef}
\end{equation}
where $\chi^2 (\sigma_{\rm neutron})$ is the chi-square of the fit for a given SD WIMP-neutron cross section $\sigma_{\rm neutron}$, and $\sigma_{\rm min}$ is the cross section which gives the minimum chi-square.
The limit $\sigma_{90}$ was obtained using the following relation:

\begin{equation}
  \frac{\int_0^{\sigma_{90}}L(\sigma_{\rm neutron})\; d\sigma_{\rm neutron} }{\int_0^{\infty}L(\sigma_{\rm neutron})\; d\sigma_{\rm neutron}}=0.9
\end{equation}

The obtained 90\% CL upper limit for a $200\; {\rm GeV}/c^2$ WIMP is $ 4.1\times 10^{-39} \;\rm cm^2$.
The fitted energy spectra of the $\beta$-depleted sample for each period of $200\; {\rm GeV}/c^2$ WIMPs (90\% CL upper-limit) are shown in Figure \ref{zoom90}. 
Step structures seen at $\rm 60\; keV_{ee}$ were induced by energy-dependent correction of $\beta$ mis-ID. 
The scaling factor for each energy region (every $\rm 10\;  keV_{ee}$) used in this correction is shown in Figure \ref{beta200}.
This scaling factor is 1.5 times larger between $\rm 60\textendash 65\;  keV_{ee}$ than between $\rm 55\textendash 60\; keV_{ee}$.
Due to the LXe purification, the activity of $^{14}$C decreased as time proceeds. The activity of $^{39}$Ar, which presumably emanates from the inner structure of the detector, was increasing.
For the check of the classification of each RI into 3 samples and the distribution of each BG spectrum, the $\beta$-depleted, $\beta$-enriched, and $^{214}$Bi samples of period 1 are also shown in Figure \ref{allsample}.

\begin{figure}[t]
\begin{center}
\includegraphics[clip,width=9.0cm]{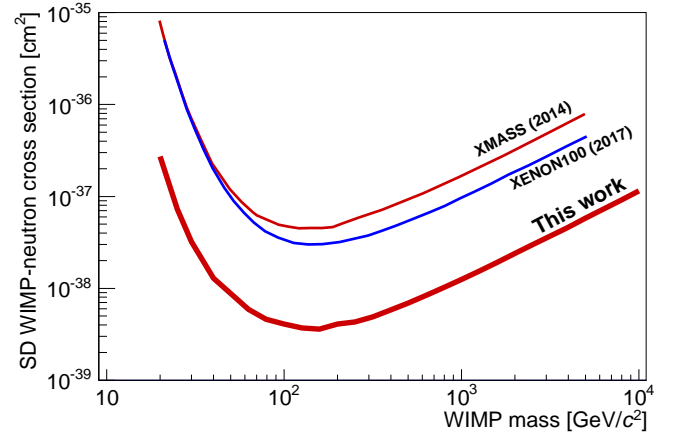}
\caption{90\% CL upper-limit for the WIMP-neutron cross section obtained by inelastic scattering searches. 
The result of this analysis is shown as a solid bold line.
The results of other experimental SD inelastic scattering searches are shown with solid lines: XMASS (2014) \cite{uchida}, XENON100 (2017) \cite{xenon100}. }
\label{limit}
\end{center}
\end{figure}

The 90\% CL upper limits for WIMPs obtained by inelastic scattering searches are shown in Figure \ref{limit}.
This result is the most stringent result to date of WIMP searches via the inelastic channel.

\section{Conclusion}

In this paper, an improved WIMP search via inelastic scattering using $\rm 327\; kg\times 800.0\; days$ data was described.
In addition to the data increase from \cite{uchida}, detailed evaluation of BG and particle identification using the decay time were introduced to discriminate inelastic scattering events from $\beta$-ray events. The obtained energy spectra were fitted with $\rm WIMP+BG$ MC spectra in the energy range from 30 to 200 $\rm keV_{ee}$. No significant signal was found. 
Therefore, the 90\% CL exclusion limits on the SD WIMP-neutron cross section were derived with the best limit of $4.1\times 10^{-39} \;\rm cm^2$ at $200\;{\rm  GeV}/c^2$. These limits are the most stringent among all current WIMP searches employing inelastic scattering.

\section*{Acknowledgements}

We gratefully acknowledge the cooperation of Kamioka Mining and Smelting Company. 
This work was supported by the Japanese Ministry of Education,
Culture, Sports, Science and Technology,
the joint research program of the Institute for Cosmic Ray Research (ICRR),
the University of Tokyo,
Grant-in-Aid for Scientific Research, 
JSPS KAKENHI Grant Number, 19GS0204, 26104004, partially
by the National Research Foundation of Korea Grant funded
by the Korean Government (NRF-2011-220-C00006),
and Institute for Basic Science (IBS-R017-G1-2018-a00).


\end{document}